\documentclass[11pt,final]{article}

\usepackage[noblocks]{authblk}

\usepackage{multirow}
\usepackage{amsmath,amssymb}
\usepackage[pdftex,final]{graphicx}
\usepackage[labelfont=bf,
            labelsep=period,
            figurename=Figure,
            justification=raggedright]
            {caption}
\usepackage[letterpaper,margin=1in]{geometry}
\usepackage{url,ragged2e,xcolor,stix} 
\usepackage[sf,bf,medium,compact]{titlesec}
\usepackage{booktabs}

\usepackage[sort&compress,super,comma]{natbib}

\let\oldthebibliography=\thebibliography
\let\oldendthebibliography=\endthebibliography
\renewenvironment{thebibliography}[2]{%
     \oldthebibliography{#1}%
     \setcounter{NAT@ctr}{#2}%
}{\oldendthebibliography}

\date{}
\title{\vspace{-1in}Stiffness of the human foot and evolution of the transverse arch}
\author[1]{Madhusudhan Venkadesan\thanks{\texttt{m.venkadesan@yale.edu}}}
\author[2,3,4]{Marcelo A.\ Dias\thanks{Present address: Department of Engineering, Aarhus University, Inge Lehmanns Gade 10, 8000 Aarhus C, Denmark}}
\author[5]{Dhiraj K.\ Singh\thanks{Present address: Engineering Mechanics Unit, Jawaharlal Nehru Centre for Advanced Scientific Research, Jakkur, Bangalore, 560064, India.}}
\author[5]{Mahesh M.\ Bandi\thanks{\texttt{bandi@oist.jp}}}
\author[6]{Shreyas Mandre\thanks{\texttt{shreyas\char`_mandre@brown.edu}}}
\affil[1]{Department of Mechanical Engineering \& Materials Science,
Yale University, New Haven, CT 06520, USA}
\affil[2]{School of Science, Aalto University, Espoo, FI-02150, Finland}
\affil[3]{Nordic Institute for Theoretical Physics (NORDITA), Roslagstullsbacken 23, SE-106 91 Stockholm, Sweden}
\affil[4]{Department of Physics and Astronomy, James Madison University, Harrisonburg, VA 22807, USA}
\affil[5]{Collective Interactions Unit, OIST Graduate University, 1919-1 Tancha, Onna-son, Okinawa 904-0495, Japan}
\affil[6]{School of Engineering, Brown University, Providence, RI 02912, USA}

\newcounter{firstbib}
\begin{document}
\maketitle
\section*{Abstract}
{
Foot stiffness underlies its mechanical function, and is central to the evolution of human bipedal locomotion.\citep{Susman1983aa,Bramble2004aa,Harcourt-Smith2004aa,DeSilva2010ab,Pontzer2012aa}
The stiff and propulsive human foot has two distinct arches, the longitudinal and transverse.\citep{Harcourt-Smith2004aa,DeSilva2010ab,Pontzer2012aa}
By contrast, the feet of non-human primates are flat and softer.\citep{Ker1987aa,Bennett1989aa,DAout2002aa}
Current understanding of foot stiffness is based on studies that focus solely on the longitudinal arch,\citep{Morton1924ab,Elftman1935aa,Hicks1954aa,Bojsen-Moller1979aa,Williams2000aa,Prang2016aa} and little is known about the mechanical function of the transverse arch.
However, common experience suggests that transverse curvature dominates the stiffness; a drooping dollar bill stiffens significantly upon curling it along the transverse direction, not the longitudinal.
We derive a normalized curvature parameter that encapsulates the geometric principle\citep{Love1927aa} underlying the transverse curvature-induced stiffness.
We show that the transverse arch accounts for almost all the difference in stiffness between human and monkey feet (vervet monkeys and pig-tailed macaques) by comparing transverse curvature-based predictions against published data on foot stiffness.\citep{Ker1987aa,Bennett1989aa}
Using this functional interpretation of the transverse arch, we trace the evolution of hominin feet\citep{Day1964aa,Pontzer2010aa,Ward2011aa,Haile-Selassie2012aa,Harcourt-Smith2015aa} and show that a human-like stiff foot likely predates {\slshape Homo} by $\sim 1.5$~million years, and appears in the $\sim 3.4$~million year old fossil from Burtele.\citep{Haile-Selassie2012aa}
A distinctly human-like transverse arch is also present in early members of {\slshape Homo}, including {\slshape Homo naledi},\citep{Harcourt-Smith2015aa} {\slshape Homo habilis},\citep{Day1964aa} and {\slshape Homo erectus}.\citep{Pontzer2010aa}
However, the $\sim 3.2$ million year old {\slshape Australopithecus afarensis}\citep{Ward2011aa} is estimated to have possessed a transitional foot, softer than humans and stiffer than other extant primates.
A foot with human-like stiffness probably evolved around the same time as other lower limb adaptations for regular bipedality,\citep{Harcourt-Smith2004aa,Raichlen2010aa,Ward2011aa,Crompton2012aa} and well before the emergence of {\slshape Homo}, the longitudinal arch, and other adaptations for endurance running.\citep{Bramble2004aa}
}

\clearpage

\begin{figure}[!hbtp]
\centering
\includegraphics{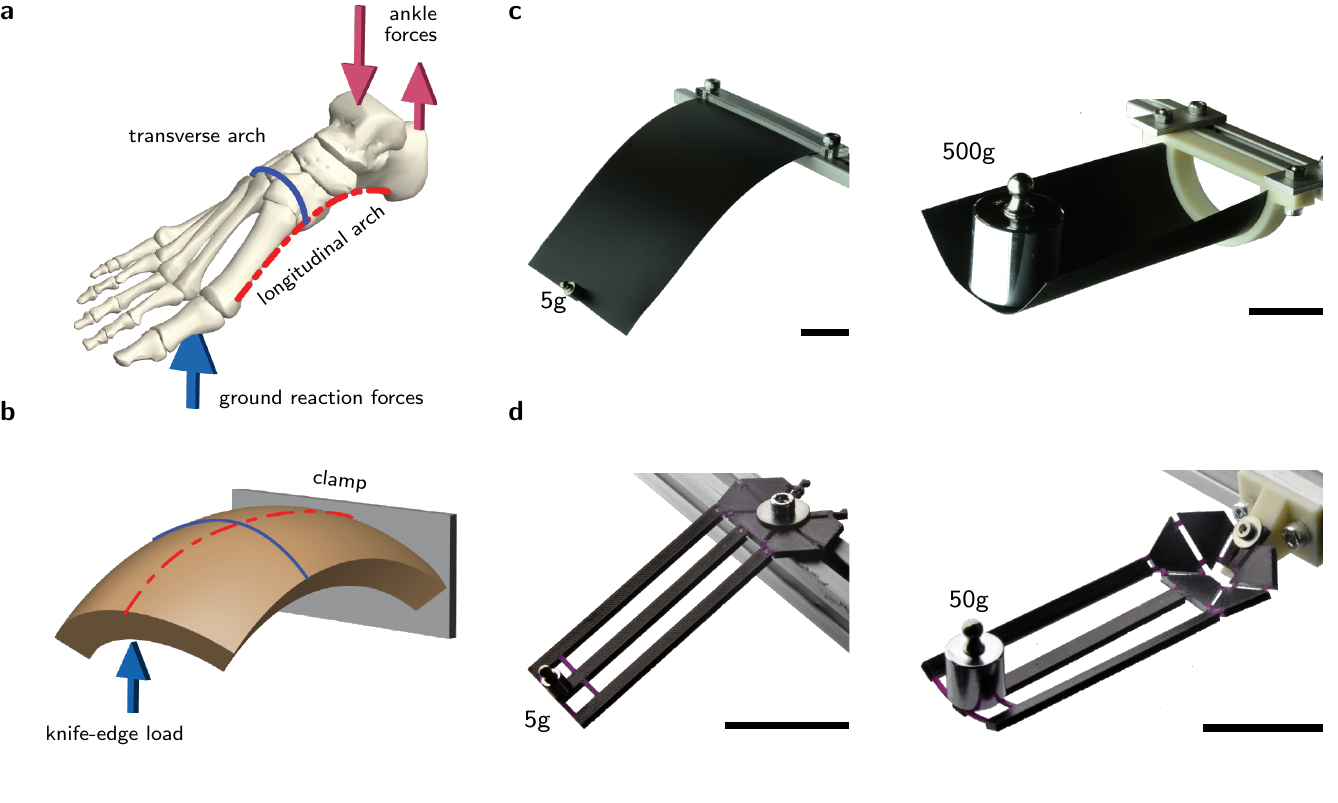}
\caption{
{\bfseries Transverse curvature in human feet, and its effect on stiffness.}
{\bfseries a,} The human foot has two distinct arches, the longitudinal and the transverse. The arched structure of the foot, and the bending loads experienced by it {\bfseries b,} are mathematically represented as an elastic shell that is clamped at the proximal end (ankle), and a distributed knife-edge load along the distal end (forefoot). {\bfseries c, d,} Illustrative demonstrations of continuum ({\bfseries c}) and discrete ({\bfseries d}) elastic structures that stiffen upon introducing transverse curvature. The scale bars are 5~cm.
}
\label{fig:relevance}
\end{figure}
\begin{figure}[!bht]
\centering
\includegraphics{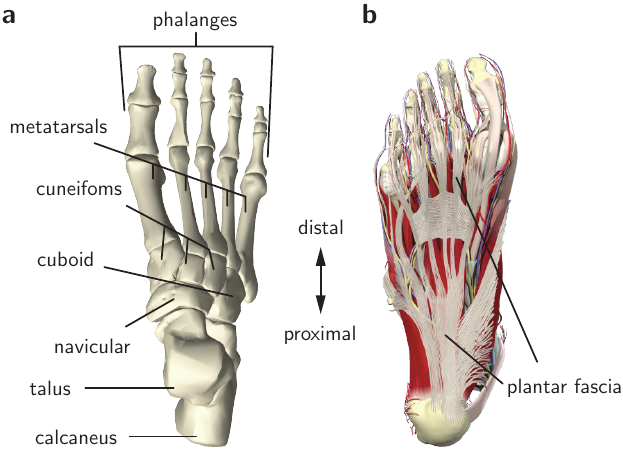}
\caption{{\bfseries Illustrated anatomy of the foot.}
{\bfseries a,} The cuneiforms, cuboid and the navicular are collectively referred to as the tarsal bones.
{\bfseries b,} The plantar fascia, a tough elastic band, extends from the calcaneus to the distal end of the phalanges. The fascia split and rejoin at multiple locations.
}
\label{fig extended: foot anatomy}
\end{figure}

The arched human midfoot acts as a stiff propulsive lever,\citep{Elftman1935aa,Bojsen-Moller1979aa,Holowka2017aa} and effectively transmits forces from the ankle to the forefoot (Fig.~\ref{fig:relevance}\ a, b).
The flat feet of non-human primates are significantly softer,\citep{DAout2002aa} and severely bend at the midfoot when the forefoot is loaded (midfoot break).\citep{Elftman1935aa,Susman1983aa,DeSilva2010ab,DAout2002aa}
Similarly, studies using 3D X-ray imaging of healthy and flatfooted humans\citep{Kido2013aa} suggest that healthy feet are over 100\% stiffer than flatfooted ones;
when subjected to body weight loading, the midfoot deformation was two-fold greater for flatfooted subjects, and registered as an increased dorsiflexion of the hallux relative to the cuneiforms.\citep{Kido2013aa}
The plantar fascia (Fig.~\ref{fig extended: foot anatomy}) and longitudinal arch are hypothesized to underlie the higher stiffness of the arched human foot.\citep{Morton1924ab,Elftman1935aa,Hicks1954aa,Bojsen-Moller1979aa,Williams2000aa,Prang2016aa}
However, cadaver studies that transect the plantar fascia show a decrease of less than 25\% in foot stiffness (table~\ref{etab:Results})\citep{Ker1987aa}.
The transverse arch emerges as an alternative source of stiffness.
Even slightly curling a thin elastic structure in the transverse direction causes it to substantially stiffen (Fig.~\ref{fig:relevance}c); a principle that is employed from the engineering design of measuring tape to the handling of pizza.
A similar effect is evident in discrete mechanical mimics of the foot (Fig.~\ref{fig:relevance} d),\citep{Venkadesan2017aa} and rayed fish fins.\citep{Nguyen2017aa}

Direct stiffness measurement using cadaveric feet (summarized in table~\ref{etab:Results}) show that human feet are nearly three-fold stiffer than that of the semi-terrestrial vervet monkey ({\it Chlorocebus aethiops}) or the pig-tailed macacque ({\it Macaca nemestrina}).
Such a comparison between human feet and other primates is however confounded by overall size differences (Fig.~\ref{fig extended:foot samples}).
For example, a curved sheet of paper may be stiffer than a flat sheet simply by virtue of difference in thicknesses, and not necessarily due to its curvature.
First consider how size affects the bending stiffness of a flat elastic plate made of a homogeneous material.
The stiffness of a cantilevered flat plate\citep{Love1927aa} of thickness $t$, width $w$, length $L$, Young's modulus $E$ and Poisson's ratio $\nu$ is $K_{\rm plate} = E w t^3/(4 L^3 (1-\nu^2))$ (appendix~\ref{suppsec:math model}).
Normalizing the stiffness of a cantilevered curved shell (Fig.~\ref{fig:relevance} b) by the stiffness of a flat plate with identical material, length, width and thickness isolates the contribution of curvature.
This normalized stiffness ($\hat{K}$) is given by,
\begin{equation}
\hat{K} = \frac{K}{K_{\rm plate}}.
\label{eqn:Khat}
\end{equation}
The ratio of the normalized stiffnesses of two curved shells of identical materials, but different lengths ($L_1$, $L_2$), widths ($w_1$, $w_2$) and thicknesses ($t_1$, $t_2$), quantifies the individual contribution of curvature differences, and is given by,
\begin{equation}
\frac{\hat{K}_1}{\hat{K}_2} = \frac{K_1}{K_2}\left(\frac{L_1}{L_2}\right)^3 \left(\frac{t_2}{t_1}\right)^3 \frac{w_2}{w_1}.
\label{eqn:stiffness ratio}
\end{equation}
By isolating the contribution of curvature, this ratio helps compare feet of different sizes by factoring the influence of overall size, under the hypothesis that the material properties are conserved.
The human foot still remains over $2.5\times$ stiffer than those of {\it C.~aethiops} and of {\it M.~nemestrina} ($\hat{K}_{h}/\hat{K}_{c}$ and $\hat{K}_{h}/\hat{K}_{m}$ in table~\ref{tab:Khat results}).
Transecting the plantar fascia softens the human foot by $\sim 23$\%, but only $\sim 4$\% for the macacque, and therefore the human foot remains over $2\times$ stiffer than the macaque ($\hat{K}_{h,{\rm pf}-}/\hat{K}_{m,{\rm pf}-}$ in table~\ref{tab:Khat results}).
The contribution of the arched structure of the human foot to its stiffness is therefore substantially greater than can be accounted for by the longitudinal arch and the plantar fascia.
How much of this difference is attributable to the transverse curvature versus differences in their material properties arising from soft tissues and muscle activity?
\begin{table}
\centering
\caption{
{\bfseries Table of estimated normalized curvature, and the estimated and predicted normalized stiffnesses.} Statistics for all estimates were obtained using Monte Carlo simulations (see methods). Normalized curvature for human feet ($\hat{c}_{h}$) are derived from published morphometric data (Fig.~\ref{fig:arch curvature} and table~\ref{etab:extant morphometrics}). Based on these, predictions are generated for the normalized stiffness of the human foot ($\hat{K}_{h,{\rm pred}}$) using equation~\eqref{eqn:Khat prediction}). Independently, published data\citep{Ker1987aa,Bennett1989aa} are used to estimate the ratio of the normalized human foot stiffness ($\hat{K}_{h}$) to that of {\it C.~aethiops} ($\hat{K}_{c}$) or {\it M.~nemestrima} ($\hat{K}_{m}$). The stiffness of the human foot with the plantar fascia transected ($\hat{K}_{h,{\rm pf}-}$) is compared against that of {\it M.~nemestrima} with the plantar fascia transected ($\hat{K}_{m,{\rm pf}-}$). Dimensional foot stiffnesses from published data\citep{Ker1987aa,Bennett1989aa} are listed in the table~\ref{etab:Results}.
}
\begin{tabular}{*3l*5c}
\toprule
Variable & mean & median & SD & 1st quartile & 3rd quartile\\
\midrule
$\hat{K}_{h}/\hat{K}_{c}$ & 2.61 & 2.38 & 1.16 & 1.80 & 3.16 \\
$\hat{K}_{h}/\hat{K}_{m}$ & 2.59 & 2.34 & 1.22 & 1.74 & 3.16 \\
$\hat{K}_{h,{\rm pf}-}/\hat{K}_{m,{\rm pf}-}$ & 2.06 & 1.87 & 0.98 & 1.39 & 2.51 \\
\midrule
$\hat{c}_{h}$ & 15.58 & 15.11 & 5.63 & 11.65 & 19.00 \\
\midrule
$\hat{K}_{h,{\rm pred}}$ & 2.04 & 1.86 & 1.09 & 1.26 & 2.62 \\
\bottomrule
\end{tabular}
\label{tab:Khat results}
\end{table}

\begin{figure}
\centering
\includegraphics{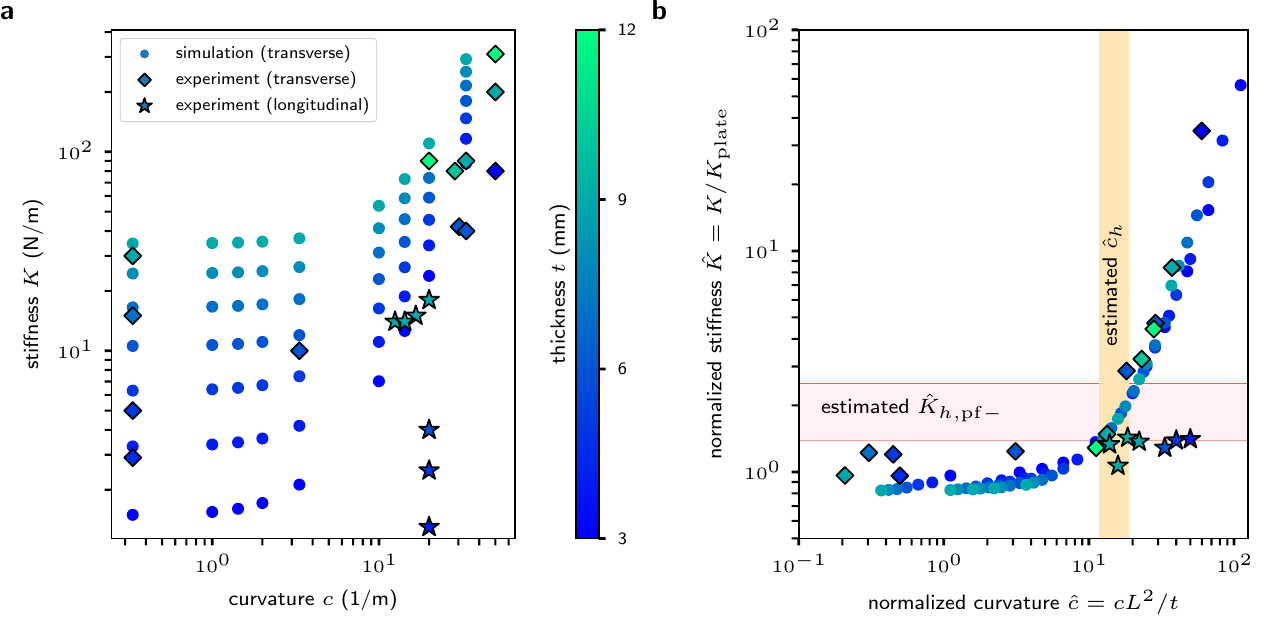}
\caption{
{\bfseries Curvature-induced stiffness.}
{\bfseries a,} Experimental measurements ($\blacklozenge$, $\star$) and numerical simulations ($\bullet$) of shells show that stiffness $K$ increases with increasing transverse curvature $c$, in addition to its dependence on the width $w$, thickness $t$, and length $L$.
{\bfseries b,} The normalized plot of $\hat{c}$ versus $\hat{K}$ shows that transverse curvature ($\blacklozenge$, $\bullet$) induces stiffening, but not longitudinal curvature ($\star$).
Overlaid on the shell data is the middle fiftieth percentile of $\hat{c}$ for human feet (vertical beige band).
Also overlaid is the middle fiftieth percentile of $\hat{K}_{h,{\rm pf}-}/\hat{K}_{m,{\rm pf}-}$ the estimated stiffness of human feet without the plantar fascia, normalized by {\it M.~nemestrina} without plantar fascia (horizontal pink band).
See table~\ref{tab:Khat results} for numerical details.
}
\label{fig:shells}
\end{figure}

We first analyze the mechanics of shells to delineate the contribution of the transverse arch.
Because transverse curvature $c$ couples longitudinal bending and in-plane stretching (supplement~\ref{suppsec:math model}), the mechanical response is governed by a balance in the elastic energy stored in bending versus stretching.
Bending and stretching energies scale differently with thickness, as $t^3$ and $t$, respectively.
Therefore, a new length-scale $\ell=\sqrt{t/c}$ emerges from this bending-stretching trade-off induced by the curvature $c$ (supplement~\ref{suppsec:subsec:Scaling analysis}).
When the shell experiences appreciable stretching, its stiffness scales as $K\sim Ewt^3/\ell^3$, unlike a flat plate that scales as $K_{\rm plate}\sim Ewt^3/L^3$ (appendix~\ref{suppsec:subsec:Scaling analysis}).
Their ratio, the normalized stiffness $\hat{K}$, depends only on the normalized curvature $\hat{c}$,
\begin{equation}
	\hat{c} = \frac{L^2}{\ell^2} = \frac{c L^2}{t}. \label{eqn:cnorm}
\end{equation}
In both experiments and computer simulations (Fig.~\ref{fig:shells}a and appendices~\ref{suppsec:math model}--\ref{suppsec:experimental shells}), the normalized stiffness $\hat{K}$ is mostly governed by $\hat{c}$ and no other parameter, as evidenced by the nearly perfect collapse onto a single master curve (Fig.~\ref{fig:shells}b).
There are two distinct regimes to the behavior of a shell; ``low curvature'' where $\hat{K}\approx 1$ is nearly independent of $\hat{c}$, and ``high curvature'' where $\hat{K}\sim\hat{c}^{3/2}$ (Fig.~\ref{fig:shells}b).
The transition between these regimes occurs around $\hat{c} = 10$.
The longitudinal arch however has no such effect on stiffness (Fig.~\ref{fig:shells}b), highlighting the dominant role of the transverse arch.
We note that continuum shell models are not an exact model of the foot (see appendix~\ref{suppsec:subsec:Validity of the shell approximation} for validity of this model).
Rather, the normalized curvature $\hat{c}$ parameterizes the strength of the coupling between out-of-plane bending and in-plane stretching, a principle that is purely geometric and independent of material differences.

\begin{figure}[!ht]
	\centering
\includegraphics{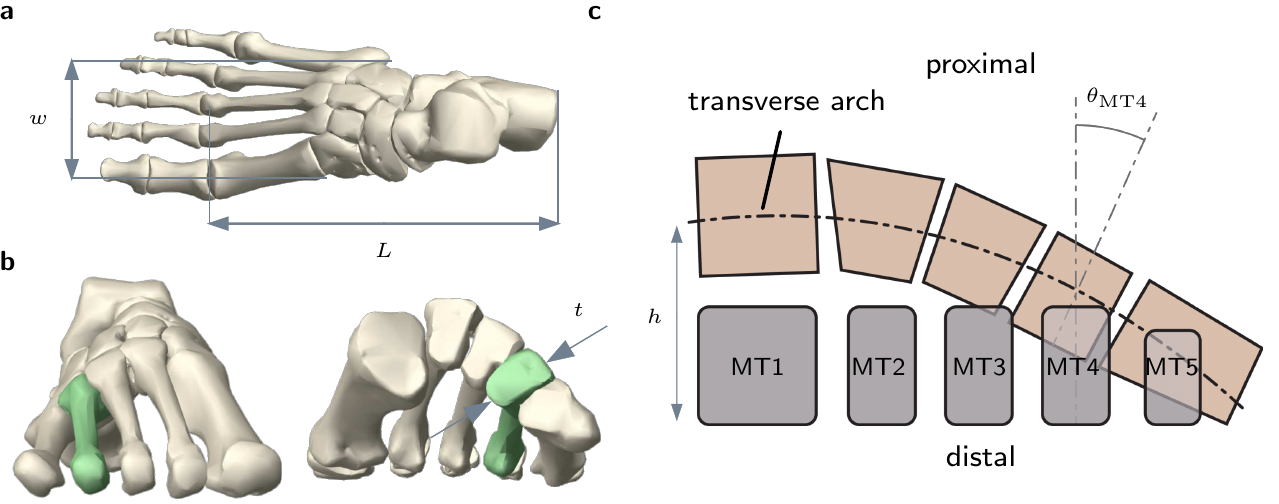}

\includegraphics{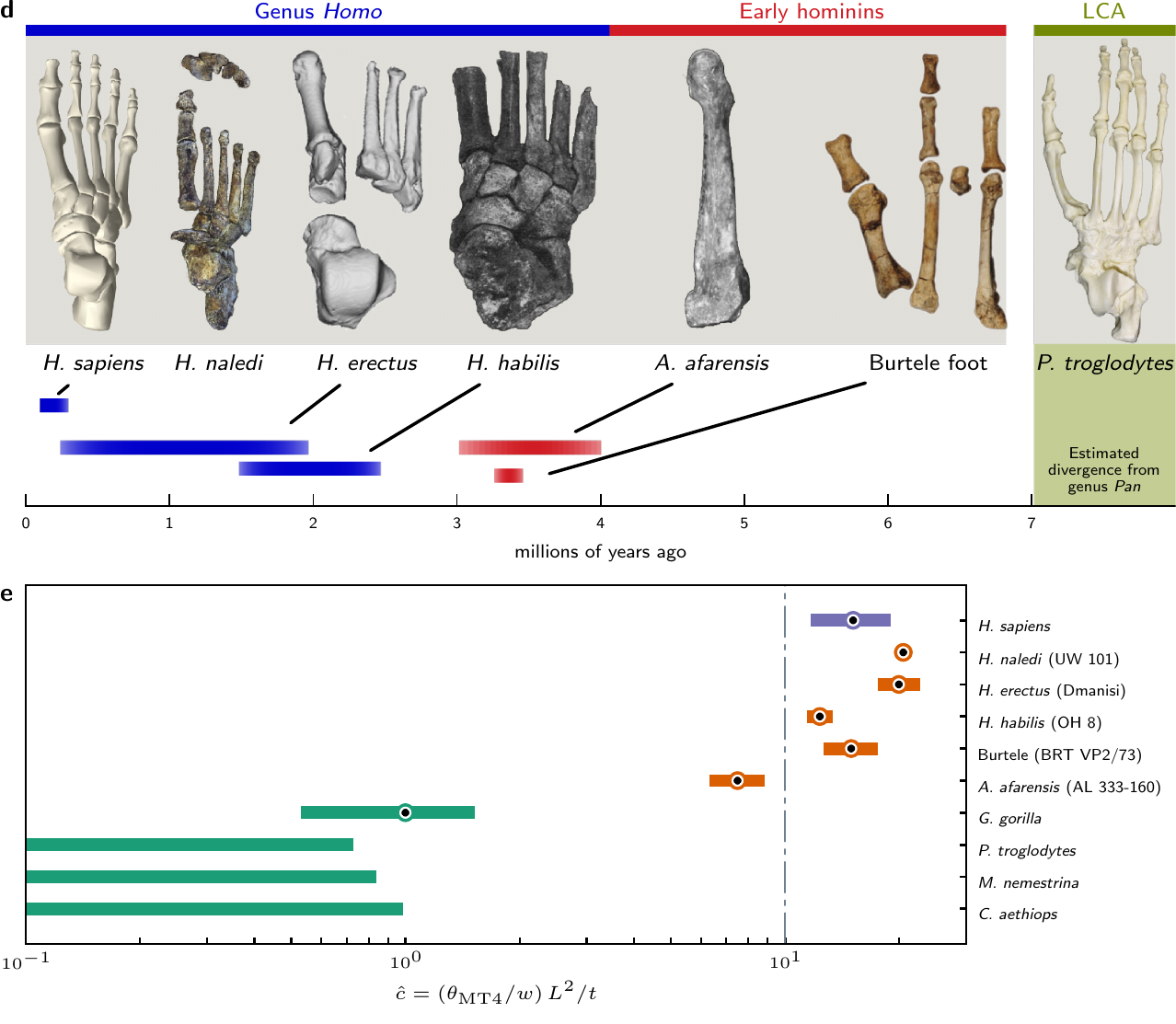}

\caption{
{\bfseries Transverse curvature of biological feet.} {\bfseries a, b,} Definitions of the length $L$, width $w$ and thickness $t$ (fourth metatarsal in green), and
{\bfseries c,} torsion of the fourth metatarsal $\theta_{\rm MT4}$ arising due to the transverse arch.
{\bfseries d,} Fossil feet used in our analyses, and the estimated timeline for the respective species. Foot of {\it Pan troglodytes} is shown in place of the last common ancestor (LCA) of humans and chimpanzees.
{\bfseries e,} Median ($\circledbullet$) and the middle $50^{\rm th}$ percentile ({\color{gray}\rule{16pt}{6pt}})
of the normalized curvature $\hat{c}$ are shown on a logarithmic scale.
The vertical dash-dot line at $\hat{c}=10$ indicates the transition from a soft plate to a stiff shell (Fig.~\ref{fig:shells}b).
Tables~\ref{etab:extant morphometrics}~and~\ref{etab:fossil morphometrics} summarize the morphometric data used in estimating $\hat{c}$.
}
\label{fig:arch curvature}
\label{fig:evolution}
\end{figure}

How well does the normalized transverse curvature predict differences in the normalized stiffness between humans and monkeys?
Estimating $\hat{c}$ requires morphometric measurements of all three dimensions of the foot (Fig.~\ref{fig:arch curvature}a, b).
However, direct measurement in fossil specimens is often impossible because of their condition (e.g.\ Fig.~\ref{fig:evolution}d).
We therefore derive a curvature estimate that relies on torsion of the fourth metatarsal $\theta_{\rm MT4}$, a widely reported measurement for both extant and fossil hominin feet,\citep{Pontzer2010aa,Ward2011aa,Haile-Selassie2012aa} and a known quantitative measure of the transverse arch.\citep{Harcourt-Smith2004aa,Pontzer2010aa,Ward2011aa,Haile-Selassie2012aa}
When the proximal metatarsal heads form a transverse arch and the distal metatarsal heads rest on the ground, the lateral metatarsals increasingly acquire torsion about their long axis (Fig.~\ref{fig:arch curvature}b,~c).
The torsion $\theta_{\rm MT4}$ arises from the curvature $c$ over the width $w$ of the tarso-metatarsal articulation, and therefore the curvature is given by $c=\theta_{\rm MT4}/w$.
Using equation~\eqref{eqn:cnorm} we find,
\begin{equation}
\hat{c}=\frac{\theta_{\rm MT4}}{\left(w/L\right)\left(t/L\right)}.
\label{eqn:torsioncurvature}
\end{equation}
Normalized transverse curvature of human feet $\hat{c}_h$ is $15.6 \pm 5.6$ (mean$\pm$standard deviation, table~\ref{tab:Khat results}).
The mean predicted normalized stiffness (equation~\ref{eqn:Khat prediction}) for this normalized transverse curvature of the human foot is $K_{h,{\rm pred}}=2.04$ (table~\ref{tab:Khat results}).
Because monkey feet are flat, we set their normalized stiffness to 1, i.e.\ the hypothesis that the feet of monkeys would be no stiffer than a perfectly flat structure that is identical in all other regards.
Under this hypothesis, experimentally estimated $\hat{K}_h = \hat{K}_h/\hat{K}_c = \hat{K}_h/\hat{K}_m \approx 2.6$.
However, this estimate of $\hat{K}$ with intact feet includes the contribution of both the arches.
To assess the individual contribution of the transverse arch, we compare stiffnesses after transecting the plantar fascia (table~\ref{etab:Results}), and find that the experimental mean of $\hat{K}_{h,{\rm pf}-}=2.06$ agrees with the predicted mean $\hat{K}_{h,{\rm pred}}=2.04$ (table~\ref{tab:Khat results} and the middle 50$^{\rm th}$ percentile in Fig.~\ref{fig:shells} b, pink band).
We conclude that the transverse arch, characterized by its normalized curvature $\hat{c}$, dominates foot stiffness.

The estimated $\hat{c}$ for extinct and extant hominin feet suggest a gradual evolution of a human-like foot stiffness (Fig.~\ref{fig:evolution}d, e).
As expected,\citep{Morton1924ab,Bennett1989aa,Pontzer2012aa} the vervet monkey, macacaque, chimpanzee and gorilla feet are substantially flatter than humans, with no appreciable transverse arch, and well below the stiffness transition at $\hat{c}\approx 10$.
Also as expected,\citep{Bramble2004aa,Bramble2004aa,Harcourt-Smith2004aa} the feet of fossils identified as genus {\it Homo} have a transverse arch that falls within normal human range, including {\it Homo naledi},\citep{Harcourt-Smith2015aa} {\it Homo habilis} (OH-8),\citep{Day1964aa} and {\it Homo erectus} (Dmanisi).\citep{Pontzer2010aa}
Surprisingly however, the transverse arch of the $\sim 3.4$ million year old Burtele foot (unidentified species)\citep{Haile-Selassie2012aa} also falls within normal human variation.
It was previously hypothesized to possess a gorilla-like foot and gait, primarily based on the proximal articulation of the hallux, and the apparent lack of a longitudinal arch.\citep{Haile-Selassie2012aa}
Also surprisingly, $\hat{c}$ of the $\sim 3.2$~million year old {\it Australopithecus afarensis} (AL-333) falls outside human range, despite the fourth metatarsal torsion being well within human range.\citep{Ward2011aa}
This is despite using a gorilla template\citep{Prang2016aa} to estimate overall foot length and width; a chimpanzee template would further separate it from humans.

Although the transverse arch underlies the ability of the foot to act as a rigid propulsive lever in walking, it does not imply all human-like capabilities such as endurance running.
By virtue of its transverse arch, the Burtele hominin was likely to have possessed a foot capable of a human-like walking gait.
The absence of a longitudinal arch (lesser elastic energy storage), long toes and an abducted hallux (stress injury risk) however indicate poor endurance running capabilities.\citep{Bramble2004aa}
The {\it A.~afarensis} foot is likely much softer than {\it Homo} and Burtele, yet slightly stiffer than the gorilla by virtue of being close to, but below the transition region of $\hat{c}\approx 10$.
It probably possessed a transitional gait; partly human-like and striding, but with greater midfoot flexibility.
This is consistent with the 3.66 million year old Laetoli~G footprints thought to have been made by {\it A.~afarensis}, which indicate partially human-like foot function.\citep{Raichlen2010aa,Crompton2012aa}
Our findings contest the current interpretation of the {\it H.~naledi}\ foot.\citep{Harcourt-Smith2015aa}
A weak or absent longitudinal arch has raised debate on the gait of {\it Homo naledi} although it has been previously noted that its metatarsal torsion is within human variation (table~\ref{etab:fossil morphometrics}), as are several other pelvic, leg and foot adaptations.\citep{Harcourt-Smith2015aa}
Based on its transverse arch,\ but a weak or absent longitudinal arch, we propose a human-like walking gait for {\it H.~naledi},\ but poor endurance running abilities.

The talotarsal joints form an elastic foundation in series with the stiff transverse arch in humans.
Therefore, foot muscles that affect structural stiffness of the inter-tarsal joints may significantly modulate foot stiffness; an essential aspect of healthy foot function in humans.\citep{Bates2013aa}
Cadaveric and {\it in vivo} evidence suggest such a role for the \emph{tibialis posterior} muscle,\citep{Kitaoka1997ab} which is phasically activated in mid-stance during walking.\citep{Maharaj2016aa}
The \emph{tibialis posterior} suddenly relaxes\citep{Maharaj2016aa} in late stance, thereby abruptly increasing foot flexibility,\citep{Holowka2017aa} and recovering the elastic energy stored in the plantar fascia during touch-down.\citep{Ker1987aa}
The transverse arch may therefore act as a mechanical clutch (activated by \emph{tibialis posterior}) that engages or disengages the elastic recoil from the longitudinal arch.
Such a role in elastic energy storage and recovery is particularly important for reducing the metabolic cost of running, and supports the hypothesis that the longitudinal arch is a later adaptation for endurance running.\citep{Bramble2004aa}
Finally, our study calls for reorienting clinical practice on foot function in walking towards the transverse arch, instead of the current focus on the longitudinal arch.

\paragraph{Acknowledgments} Funding support from the Human Frontier Science Program. Access to skeletal specimens were provided by Gary Aronsen, Kristof Zyskowski, Eric Sargis, Yale Biological Anthropology Labs and the Yale Peabody Museum. K.J.~Meacham~III\ provided experimental support.
\paragraph{Author contributions} M.V.,\ M.M.B.,\ and S.M.\ conceived of the study; M.A.D.\ and S.M.\ performed the mathematical modeling in consultation with M.V.; D.K.S.\ and M.M.B.\ performed the shell experiments; M.V.\ performed the biological data collection and analyses; M.V.\ wrote the main paper, S.M.,\ M.A.D.,\ M.M.B.,\ and M.V.\ wrote the supplement, and all authors jointly edited the entire paper.

\renewcommand{\refname}{References}

\clearpage
\renewcommand{\theequation}{M.\arabic{equation}}
\section*{Methods}
\subsection*{Numerical simulations}
We simulated the elastic response of arched shells using the \emph{Shell} interface in the \emph{3D Structural Mechanics} module of COMSOL Multiphysics v5.1\ (COMSOL AB, Stockholm, Sweden).
The transverse arch is represented by the map for the central plane of the shell given by
\begin{equation}
	\label{eq:Geo_T}
		\mathbf{S}_{\mathrm{T}}(x,y)=\left(x,R_{\mathrm{T}}\sin \theta_y ,R_{\mathrm{T}}\cos \theta_y \right)
\end{equation}
where $\theta_y = y/{R_{\mathrm{T}}}$, and the longitudinal by
\begin{equation}
	\label{eq:Geo_L}
		\mathbf{S}_{\mathrm{L}}(x,y)=\left(R_{\mathrm{L}}\sin \theta_x,y,R_{\mathrm{L}}\cos \theta_x\right)
\end{equation}
where $\theta_x = {x}/{R_{\mathrm{L}}}$, and in both cases $x\in\left[-L/2,L/2\right]$, $y\in\left[-w/2 ,w/2 \right]$.
For all the simulations, we set $L=0.1$~m and $w=0.05$~m. 
The material was modeled as linearly elastic with Young's modulus $E=3.5$~MPa, Poisson's ratio $\nu=0.49$, and mass density $\rho=965$~kg/m$^3$ to match the material used in the experiments.

\begin{figure}
\centering
\includegraphics[width=\textwidth]{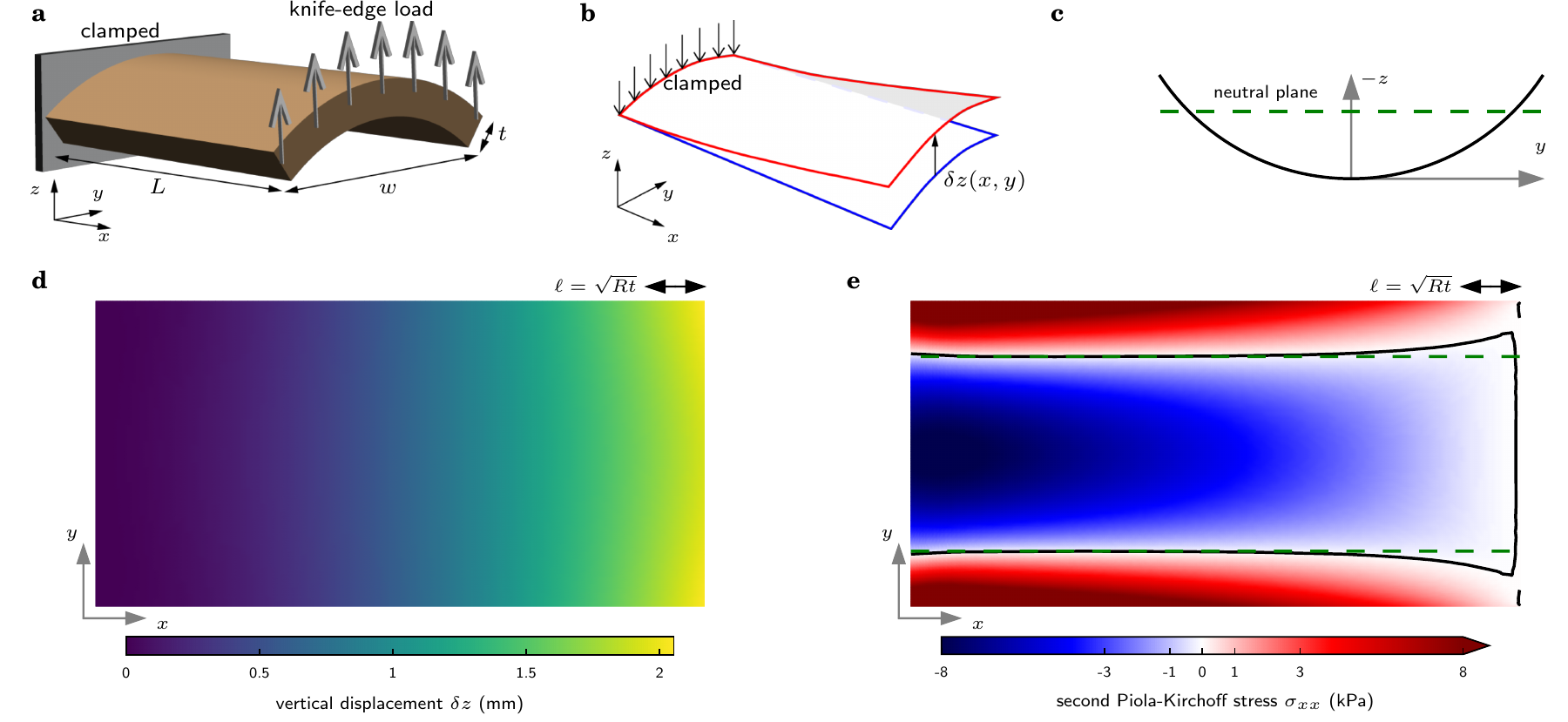}
\caption{Mathematical and computational analysis of continuum elastic shells. {\bf a.} The shell is clamped on one end, and loaded with a knife-edge on the other. It is of length $L$, width $w$, thickness $t$, and has radius of curvature $R$ (curvature $c=1/R$). {\bf b.} The free end displaces by a height $\delta z$ upon loading, and reaction forces at the clamped end resist deformation. {\bf c.} A cross-sectional view of the shell shows the location of the neutral plane, if the shell were to act as an elastic beam. {\bf d.} Out-of-plane ($z$-axis) displacement profile for one numerical simulation of a shell ($L=0.1$~m, $w=0.05$~m, $t=0.003$~m, $R=0.03$~m). Most of the displacement happens close to the loaded edge, unlike an elastic beam. {\bf e.} The stress component $\sigma_{xx}$ is shown as a color-map of the undeformed shell. In an elastic beam, the intersection of the neutral plane with the shell (panel c) would exactly match the locations of zero stress. Because of curvature-induced in-plane stretching, the zero stress curve differs from the neutral plane predictions in the vicinity of the loaded edge, and to a lesser extent, near the clamped boundary.}
\label{fig extended:numerics}
\end{figure}

The boundary at $x=-L/2$ is clamped, i.e.\ zero displacements and rotations.
The conditions at the other boundary $x=L/2$, are a uniform shear load $\mathcal{T}$, zero bending moment along $z$, and zero in-plane traction so that the displacements are free (see Figure \ref{fig extended:numerics} for axes orientations).

We solve this model for a range of thicknesses $t$, from 3 to 9~mm in steps of 1~mm, and transverse curvature radii $R_{\mathrm{T}} = 0.03,~0.05,~0.07,~0.1,~0.3,~0.5,~0.7,~1,~3$m. For each combination of $t$ and $R_{\mathrm{T}}$, shear $\mathcal{T}$ ranging from 0 to 1 N/m is applied in increments of 0.1 N/m. 
The resulting out-of-plane displacement $\delta z$ is measured (Fig.~\ref{fig extended:numerics}(b)), and plotted against $\mathcal{T}$. The slope of these curves extrapolated to $\mathcal{T}=0$ yield the stiffness defined as $k\equiv w\,\mathcal{T}/\delta z$.
The above process is repeated for geometry in Eq.~\eqref{eq:Geo_L}.

\subsection*{Mechanical experiments}
We fabricated and measured the stiffness of shells with a transverse arch or a longitudinal arch, and of a flat plate.
These were all fabricated using polymer moulding techniques with PDMS (Poly dimethyl siloxane).
The mould was fabricated using additive manufacturing (3D printed using ProJet 460Plus, 3D Systems).
The printed mould was few millimeters in thickness, with one side left open.
PDMS silicone elastomer (Sylgard 184, Dow Corning) was employed to cast the arch in the mould.
Because the volume ratio of the base polymer to the curing agent controls the material bulk modulus for PDMS, the same ratio of 5 parts base polymer to 1 part of curing agent by weight was consistently maintained across all fabricated arches (appendix~\ref{suppsec:experimental shells}).
During an experiment, the fabricated arch was mounted on the experimental rig with help of clamps that were custom fabricated to exactly match the arch curvature.
The clamps were additively manufactured (Stratasys Dimension 1200es) with ABSPlus (Acrylonitrile butadiene styrene) thermoplastic material (glass transition temperature 108$^{\circ}$C).
One end of the clamped arch was affixed to a rigid plate attached to a free-moving horizontal translation stage to maintain a nearly zero horizontal load during the test.
The other side of the clamped arch was connected to a force sensor, which was affixed to a vertical translation stage.
The forces were measured using a data acquisition system (LabView, National Instruments) at 2 KHz for 1 second duration.
The load test was performed under quasi-static loading of the arch sample by providing small displacements (quasi-static steps) of $5 \times 10^{-5}$ m ($50~\mu$m) per step for a total of 10 quasi-static steps ($5 \times 10^{-4}$ m or $500~\mu$m).
Forces were measured after each quasi-static displacement.
The slope of the force-displacement curve is the bending stiffness $K$ for the arch sample.
Three experimental runs were conducted for each arch and their force-displacement curves were reproducible to within measurement error.

\subsection*{Monte Carlo simulations}
Anatomical variability in the size of feet (table~\ref{etab:extant morphometrics}) is incorporated using Monte Carlo simulations to generate statistics for normalized stiffnesses and curvatures (table~\ref{tab:Khat results}). The histograms generated from the Monte Carlo simulations are non-Gaussian. Therefore, the median and quartiles are reported in addition to the mean and standard deviation (SD). We used 1 million random combinations of the anatomical dimensions, where each dimension was drawn from an independent Gaussian distribution with means and standard deviations according to tables~\ref{etab:extant morphometrics}~and~\ref{etab:fossil morphometrics}. Increasing the size of the Monte Carlo beyond a million samples had no effect on the statistics of the estimated quantities, for the number of significant digits reported. Our simulations likely overestimate the variance of relevant ratios such as $w/L$ and $t/L$ in comparison to biological feet, because we do allow for independent variation of all dimensions and do not incorporate covariation that may exist.

\subsection*{Foot stiffness}
Published force versus displacement measurements of cadaveric human feet,\citep{Ker1987aa} and from monkeys ({\it Chlorocebus aethiops} and {\it Macaca nemestrina})\citep{Bennett1989aa} were used to estimate their stiffness. These published measurements\citep{Ker1987aa,Bennett1989aa} mimic the numerical and experimental design of stiffness measurements; the heel is clamped, the ball of the foot is vertically displaced with no horizontal forces, and the vertical reaction force is measured. The ratio of total measured force to the total applied displacement is the foot's stiffness $K$. The published plots, figure~3 in Ker {\it et al.}\citep{Ker1987aa}\ and figures~1--2 in Bennett {\it et al.}\citep{Bennett1989aa}, were digitized to extract values (using B.~Tummers,\ DataThief III. 2006 $<$\url{http://datathief.org/}$>$).

Normalization of the dimensional stiffness estimates of the human foot are carried out by considering the ratio of human to monkey foot stiffness. We report the normalized stiffness relative to both {\it C.~aethiops} and {\it M.~nemestrina} in table~\ref{etab:Results} according to equation~\eqref{eqn:stiffness ratio}. In addition to the ratio of stiffnesses of intact feet, we also estimate the ratio for feet where the plantar fascia was transected. This allows the quantification of the transverse arch's individual contribution to foot stiffness.

\begin{table}
\centering
\caption{Table of estimated stiffnesses from published load versus displacement data for humans\citep{Ker1987aa}, and for {\it C.~aethiops} and {\it M.~nemestrima}.\citep{Bennett1989aa} In addition to stiffness of the intact foot (shown here as $K_{\square}$) for the three species, we use stiffness of feet with the plantar fascia transected (shown here as $K_{\square, {\rm pf}-}$). These estimates were obtained by digitizing the published plots\citep{Ker1987aa,Bennett1989aa} of load versus displacement.}
\begin{tabular}{*3lc}
\toprule
Species & Foot condition & Variable & Estimated value \\
\midrule
{\it Homo sapiens} & intact & $K_{h}$ (N/mm) & 481 \\
{\it Homo sapiens} & plantar fascia removed & $K_{h,{\rm pf}-}$ (N/mm) & 369 \\
{\it Chlorocebus aethiops} & intact & $K_{c}$ (N/mm) & 132 \\
{\it Macaca nemestrina} & intact & $K_{m}$ (N/mm) & 165 \\
{\it Macaca nemestrina} & plantar fascia removed & $K_{m,{\rm pf}-}$ (N/mm) & 159 \\
\bottomrule
\end{tabular}
\label{etab:Results}
\end{table}

\subsection*{Curvature-based stiffness prediction}
The normalized stiffness $\hat{K}$ scales as the $3/2$ power of $\hat{c}$ for continuum shells for high curvature, but is independent of $\hat{c}$ and equal to 1 for low curvature. The equation to predict $\hat{K}$ based on these asymptotic behaviors is given by,
\begin{equation}
	\log\hat{K}_{\rm pred} = 
	\begin{cases}
	0, &\hat{c} < 10,\\
	\frac{3}{2}\log\left(\frac{\hat{c}}{10}\right), &\hat{c}>10.
	\end{cases}
	\label{eqn:Khat prediction}
\end{equation}

\subsection*{Feet of extant species}
Morphometrics for all species were obtained from the published literature. Additionally, for the feet of humans, {\it C.~aethiops}, and {\it M.~nemestrina}, we carried out primary measurements using specimens that were most similar in their lever length $L$ (Fig.~\ref{fig:arch curvature}) to the mean value reported in the literature. We carried out these measurements using software-based photogrammetry\citep{Schindelin2012aa} of high resolution images, and cross-verified with measurements using a digital caliper (0.01~mm resolution). The human specimen is from a 72 year old Caucasian male (Skulls Unlimited, \#11177), the {\it C.~aethiops} foot from the Yale Biological Anthropology Laboratory (YBL.3032a) and the {\it M.~nemestrina} specimen from the Yale Peabody Museum (YPM MAM 9621). These samples were used to verify that the ratios $w/L$ and $t/L$ obtained from the literature are indeed representative of a sample of similar size.

The lever length $L$ follows standard definition to be the distance from the posterior end of the calcaneus to the anterior end of the third metatarsal's distal head. The width $w$ is measured at the tarsometatarsal joint, from the most medial aspect of the distal articular surface of the navicular to the most lateral aspect of the distal articular surface of the cuboid. The thickness $t$ is defined as the dorso-plantar thickness of the proximal head of the third metatarsal, or the average of the second and fourth, when the third metatarsal data are unavailable.

\begin{table}
\centering
\caption{Foot morphometrics for extant species are modeled as Gaussian random variables. Mean values and standard deviations were obtained from reported values in the literature (methods for details). The important ratios ($w/L$ and $t/L$) were verified to be accurate using measurements of representative samples. These Gaussian random variables are used in the Monte Carlo simulations. The morphometric variables used are the lever-length of the foot $L$, width of the tarso-metatarsal articular region $w$, dorso-plantar thickness of the third metatarsal $t$, and torsion of the fourth metatarsal $\theta_{\rm MT4}$).}
\begin{tabular}{l*8c}
\toprule
Species & \multicolumn{2}{c}{$L$ (mm)} & \multicolumn{2}{c}{$w$ (mm)} & \multicolumn{2}{c}{$t$ (mm)} & \multicolumn{2}{c}{$\theta_{\rm MT4}$ (deg)}\\
{} & mean & SD & mean & SD & mean & SD & mean & SD\\
\midrule
{\it Homo sapiens} & 200 & 14.0 & 60.0 & 5.4 & 18.0 & 1.6 & 23.6 & 7.1\\
{\it Chlorocebus aethiops} & 85.0 & 4.3 & 24.0 & 1.2 & 9.0 & 0.45 & 0 & 2.5\\
{\it Macaca nemestrina} & 100 & 6.0 & 35.0 & 2.1 & 10.0 & 0.6 & 0 & 2.5\\
{\it Pan troglodytes} & 130 & 13.0 & 52.0 & 5.2 & 13.0 & 1.3 & 0 & 2.5\\
{\it Gorilla gorilla} & 176 & 17.6 & 72.5 & 7.3 & 16.0 & 1.6 & 2.2 & 1.5\\
\bottomrule
\end{tabular}
\label{etab:extant morphometrics}
\end{table}

\begin{figure}
\centering
\includegraphics{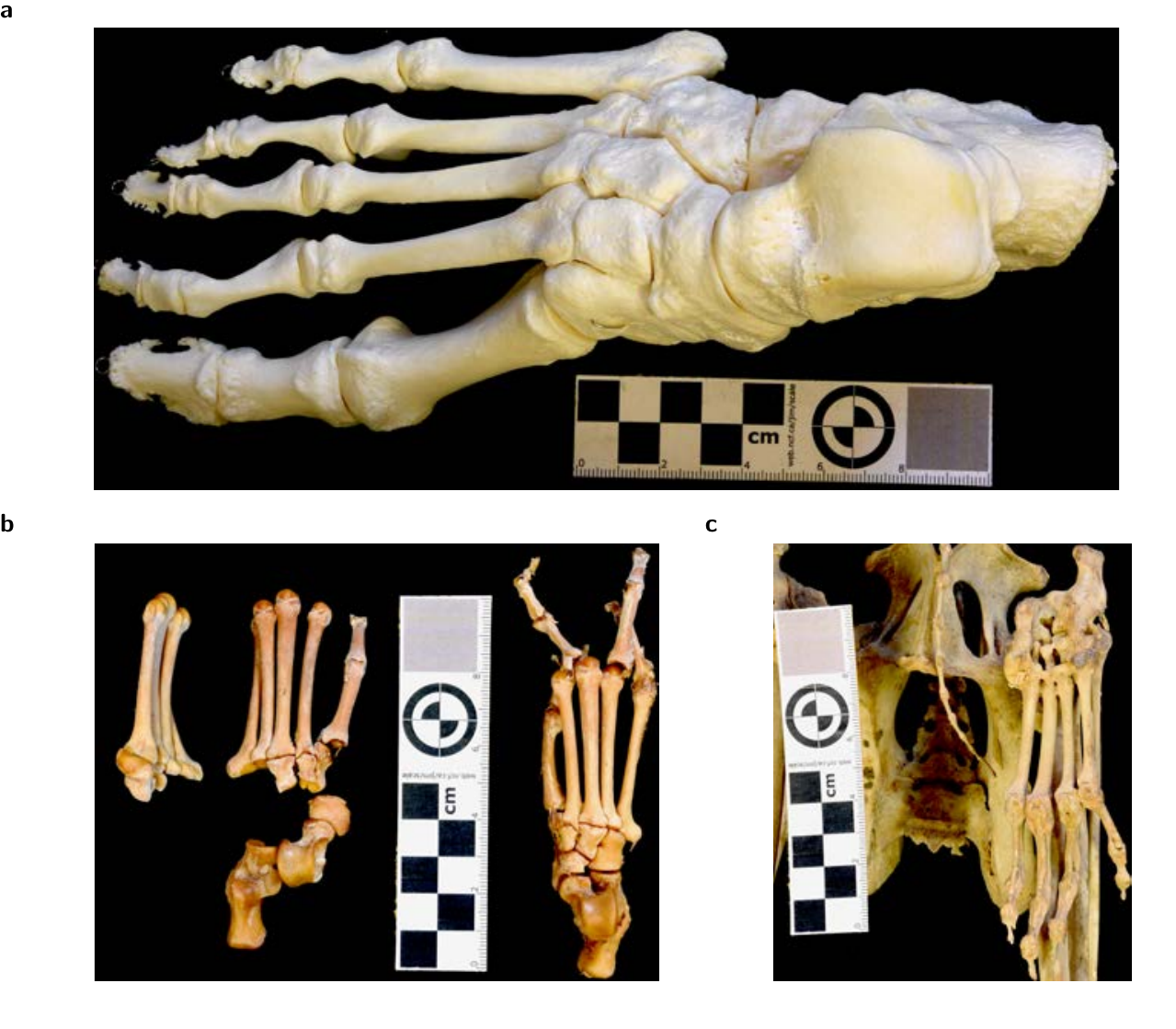}
\caption{Foot skeletons of {\bf a.} {\it Homo sapiens} (Yale Biomechanics and Control Laboratory), {\bf b.} {\it Chlorocebus~aethiops} (Yale Biological Anthropology Laboratory, YBL.3032a), and {\bf c.} {\it Macaca~nemestrina} (Yale Peabody Museum, YPM MAM 9621). The composite image for {\it C.~aethiops} shows a dorsal view of both feet, and a side view of the partly disarticulated left foot. From all specimens available, the lever length $L$ for these representative specimen were the closest to the mean $L$ reported in the literature, for the respective species.}
\label{fig extended:foot samples}
\end{figure}

Mean and standard deviation (SD) of the lever length $L$ were estimated from published data for humans,\citep{Hawes1994aa,Wang2004aa,Witana2006aa,Schultz1963aa,Gomberg1981aa}  chimpanzees,\citep{Wang2004aa,Schultz1963aa,Gomberg1981aa} gorillas,\citep{Wang2004aa,Schultz1963aa}, {\it C.~aethiops},\citep{Anapol2005aa,Schultz1963aa} and {\it M.~nemestrina}.\citep{Hamada1994aa,Sirianni1975aa,Schultz1963aa} Mean $w$ is based on 3D anatomical data from BodyParts3D for humans,\citep{Mitsuhashi2009aa} estimated from reported $w/L$ or dorsal skeletal views for chimpanzees and gorillas,\citep{Schultz1963aa,Gomberg1981aa}, and primary measurements for {\it C.~aethiops} and {\it M.~nemestrina} (Fig.~\ref{fig extended:foot samples}). To estimate the SD of $w$, we used reported variability in the medio-lateral width of the proximal metatarsal heads for all species\citep{Pontzer2010aa,Ward2011aa,Haile-Selassie2012aa} to estimate the coefficient of variation (SD/mean), and applied that to $w$. The mean and SD of $t$ were all obtained from published values,\citep{Ward2011aa,Haile-Selassie2012aa} and confirmed with primary measurements for available specimens. Torsion of the fourth metatarsal $\theta_{\rm MT4}$ is used to estimate the transverse curvature, and published values were used for all species included in this study \citep{Pontzer2010aa,Ward2011aa,Haile-Selassie2012aa,Harcourt-Smith2015aa}. For species where the feet are regarded as flat, we set $\theta_{\rm MT4}=0$~deg.

\subsection*{Fossil feet}
We use photogrammetry\citep{Schindelin2012aa} on published images of fossil feet (Fig.~\ref{fig:evolution} d), and also data tables that accompanied the publication of these fossil data to estimate necessary dimensions and ratios.\citep{Day1964aa,Pontzer2010aa,Harcourt-Smith2015aa,Ward2011aa,Haile-Selassie2012aa}

\begin{table}
\centering
\caption{Values of $L$, $w$, $t$ and $\theta_{\rm MT4}$ used in estimating the normalized curvature $\hat{c}$ of fossil samples. Variable names with a subscript $h$ refer to human values (e.g.\ $t_h$), subscript $p$ to the chimpanzee (e.g.\ $w_p$), and subscript $g$ to the gorilla (e.g.\ $L_g$). These values are represented by Normal distributions as shown in table~\ref{etab:extant morphometrics}. Variables in angled brackets, such as $\langle t_h\rangle$, refer to the mean value shown in table~\ref{etab:extant morphometrics}. See methods for details of source materials.}
\begin{tabular}{ll*6c}
\toprule
Species & Specimen & $\theta_{MT4}$ (deg) & $L$ (mm) & $w$ (mm) & $t$ (mm) & $w/L$ & $t/L$\\
\midrule
{\it H.\ naledi} & UW 101-1456 & 38.0 & 137.0 & 38.0 & 16.0 & 0.277 & 0.117 \\
{\it H.\ erectus} & D2669, D4165 & 28.0, 29.0 & -- & -- & 17.0 & $\frac{w_h}{L_h}$ & $\frac{t}{\langle t_h\rangle} \frac{t_h}{L_h}$ \\
{\it H.\ habilis} & OH 8 & 25.0 & 112 & 44.0 & -- & 0.393 & $\frac{t_h}{L_h}$ \\
Burtele & BRT VP2/73 & 26.5 & -- & -- & 13.3 & $\frac{w_g}{L_g}$ & $\frac{t}{\langle t_g\rangle} \frac{t_g}{L_g}$ \\
{\it A.\ afarensis} & AL 333-160 & 17.0 & -- & -- & 17.1 & $\frac{w_g}{L_g}$ & $\frac{t}{\langle t_g\rangle} \frac{t_g}{L_g}$ \\
\bottomrule
\end{tabular}
\label{etab:fossil morphometrics}
\end{table}

Among the fossil feet, all but the foot of {\it Homo naledi}\citep{Harcourt-Smith2015aa} were incomplete in some regard. For those incomplete feet, an extant species was selected as a template by taking into consideration published analyses of other postcranial and cranial elements. Based on this, {\it Homo sapiens} was chosen as the template for {\it Homo~erectus} (Dmanisi)\citep{Pontzer2010aa} and {\it Homo~erectus} (Olduvai hominin),\citep{Day1964aa} and {\it Gorilla gorilla} was chosen as the template for {\it Australopithecus afarensis} (AL 333)\citep{Ward2011aa} and the unknown hominin foot found in Burtele.\citep{Haile-Selassie2012aa} For example, the sole fourth metatarsal of the {\it A.~afarensis} does not permit the direct estimation of $w$. However, only the ratio $w/L$ is necessary for the analyses, and the ratio for the gorilla is used for the Monte Carlo analysis of the fossil. The metatarsal however provides a direct measurement of $t$, but not of $L$. Therefore, to estimate the ratio $t/L$, we incorporate the measured thickness $t$ and the gorilla's ratio $t_g/L_g$ by using the formula,
\begin{equation}
	\frac{t}{L} = \frac{t}{\langle t_g\rangle}\frac{t_g}{L_g},
\end{equation}
where $\langle t_g\rangle$ is the mean $t$ for the gorilla. This template-based estimation therefore incorporates direct measurements where available without assuming that the fossil exactly resembles the extant template.

\subsection*{Curvature of biological feet}
Transverse curvature $c$ is estimated using the torsion of the fourth metatarsal\citep{Pontzer2010aa} $\theta_{\rm MT4}$ as given by, $c=\theta_{\rm MT4}/w$, where $w$ is the width of the foot as measured at the tarso-metatarsal joints.
Equation~\eqref{eqn:torsioncurvature} for estimating the normalized curvature $\hat{c}$ follows from equation~\eqref{eqn:cnorm}.

\renewcommand{\refname}{References for methods}

\clearpage
\appendix
\renewcommand{\theequation}{\Alph{section}.\arabic{equation}}

\begin{center}
{\sffamily\bfseries\LARGE Appendix}
\end{center}
\vspace{0.25in}
\section{Mathematical modeling}
\label{suppsec:math model}
While detailed treatments on thin shell theory can be found in standard texts~\cite{Koiter:1966,Koiter:1970,Fluegge:1973,Niordson:1985}, we present an abridged version here to define precisely the variables that follow. We use the resulting partial differential equations to derive the scaling relation that describe the essential features of the elastic response.

\subsection{Continuum analysis of thin shells}
Let $\bar{\mathcal{S}}$ represent  a collection of material points in the reference (undeformed) state of the mid-surface of the shell. These material points in $\bar{\mathcal{S}}$ are mapped in 3D Euclidian space by a three component vector function $\mathbf{S}:\bar{\mathcal{S}}\longrightarrow\mathbb{R}^3$, where $\mathbf{S}(x^1,x^2)=\left(S_1(x^1,x^2),S_2(x^1,x^2),S_3(x^1,x^2)\right)$ and $\left\{x^1,x^2\right\}$ are the local Lagrangian coordinates on $\bar{\mathcal{S}}$. Therefore, an induced prescribed metric for the surface $\bar{\mathcal{S}}$ is calculated to be\footnote{Throughout this appendix we adopt the Einstein convention, where summation is implicit for repetitive indices when multiplying components of vectors and tensors.} $\bar{\mathbf{a}}
=\left(\partial_\alpha\mathbf{S}\cdot\partial_\beta\mathbf{S}\right)\mathbf{\mathbf{d}}x^\alpha\otimes\mathbf{\mathbf{d}}x^\beta$. The deformation from $\bar{\mathcal{S}}$ to a target (deformed) mid-surface $\mathcal{S}$ is given by a deformation map $\mathbf{F}:\bar{\mathcal{S}}\longrightarrow\mathcal{S}$, thus inducing target metric given by $\mathbf{a}
=\left(\partial_\alpha\mathbf{F}\cdot\partial_\beta\mathbf{F}\right)\mathbf{\mathbf{d}}x^\alpha\otimes\mathbf{\mathbf{d}}x^\beta$. Therefore, the reference and target components of the metric tensors, also known as \emph{first fundamental forms}, are respectively given by 
\begin{subequations}
	\label{eq:FFF}
		\begin{align}
			\label{eq:FFF1}
				\bar{a}_{\alpha\beta}&= \partial_\alpha\mathbf{S}\cdot\partial_\beta\mathbf{S}\\
			\label{eq:FFF2}
				a_{\alpha\beta}&= \partial_\alpha\mathbf{F}\cdot\partial_\beta\mathbf{F}.
		\end{align}
\end{subequations}
The set of cartesian vectors $\partial_1\mathbf{F}$ and $\partial_2\mathbf{F}$ ($\partial_1\mathbf{S}$ and $\partial_2\mathbf{S}$) span the tangent space to the mid-surface of the shell $\mathcal{S}$ ($\bar{\mathcal{S}}$). Thus, we may define the unit normals to the target $\mathcal{S}$ and reference $\bar{\mathcal{S}}$ surfaces, respectively, by $\mathbf{N}=\left(\det\mathbf{a}\right)^{-1/2}\partial_1\mathbf{F}\times\partial_2\mathbf{F}$ and $\bar{\mathbf{N}}=\left(\det\bar{\mathbf{a}}\right)^{-1/2}\partial_1\mathbf{S}\times\partial_2\mathbf{S}$. The second derivative of the mapping of the material points of both reference and target surfaces, projected onto the direction of their respective normal fields, yields the curvature tensor, or \emph{second fundamental form}. Their components, respectively for reference and target shells, are given by
\begin{subequations}
	\label{eq:SFF}
		\begin{align}
			\label{eq:SFF1}
				\bar{b}_{\alpha\beta}&= \bar{\mathbf{N}}\cdot\partial_\alpha\partial_\beta\mathbf{S}\\
			\label{eq:SFF2}
				b_{\alpha\beta}&= \mathbf{N}\cdot\partial_\alpha\partial_\beta\mathbf{F}.
		\end{align}
\end{subequations}

We define three components of the displacement field,  pure in-plane $u^\alpha(x^1,x^2)$ and out-of-plane $v(x^1,x^2)$ displacements, and they give us complete information of the deformation of the mid-surface of the shell through the function $\mathbf{F}=\mathbf{S}+\mathbf{U}$, where $\mathbf{U}(x^1,x^2)=u^\alpha(x^1,x^2)\partial_\alpha\mathbf{S}+v(x^1,x^2)\bar{\mathbf{N}}$ is the displacement vector in normal coordinates. In order to relate stresses and moments on the shell with the displacement of material points, we define the two local measures of deformation, \emph{in-plane strain}, $\varepsilon_{\alpha\beta}$, as a measure stretching and \emph{curvature strain}, $\kappa_{\alpha\beta}$, as a measure of bending. Their general definitions are 
\begin{subequations}
	\label{eq:DefMeasure}
		\begin{align}
			\label{eq:Strain}
				\varepsilon_{\alpha\beta}&\equiv \frac{1}{2}\left(a_{\alpha\beta}-\bar{a}_{\alpha\beta}\right)\\
			\label{eq:Curvature}
				\kappa_{\alpha\beta}&\equiv b_{\alpha\beta}-\bar{b}_{\alpha\beta}.
		\end{align}
\end{subequations}
For small displacements, Eqs.~\eqref{eq:DefMeasure} become
\begin{subequations}
	\label{eq:DefMeasureSmall}
		\begin{align}
			\label{eq:StrainSmall}
				\varepsilon_{\alpha\beta}&\approx \frac{1}{2}\left(\bar{\nabla}_\alpha u_\beta+\bar{\nabla}_\beta u_\alpha\right)-\bar{b}_{\alpha\beta} v \\
			\label{eq:CurvatureSmall}
				\kappa_{\alpha\beta}&\approx \bar{\nabla}_\alpha\bar{\nabla}_\beta v,
		\end{align}
\end{subequations}
where $\bar{\nabla}$ is the covariant derivative with respect the reference shell~\cite{Carmo:1976aa}. 

Once the kinematic variables have been defined in Eq.~\eqref{eq:DefMeasure}, we may write the total elastic energy as a surface integral of an energy density which is entirely expressed as a function of material constants and a linear combination of the first two invariants of the variables in Eq.~\eqref{eq:DefMeasure}\footnote{Here, we shall only consider linear constitutive laws, which justifies dropping terms that go with the third invariants $\mathrm{I\!I\!I}_{\boldsymbol\varepsilon}=\mathrm{det}\left(\boldsymbol\varepsilon\right)$ and $\mathrm{I\!I\!I}_{\boldsymbol\kappa}=\mathrm{det}\left(\boldsymbol\kappa\right)$.}, namely $\mathrm{I}_{\boldsymbol\varepsilon}=\mathrm{tr}\left(\boldsymbol\varepsilon\right)$, $\mathrm{I\!I}_{\boldsymbol\varepsilon}=\left(\left(\mathrm{tr}\boldsymbol\varepsilon\right)^2-\mathrm{tr}\left(\boldsymbol\varepsilon\cdot\boldsymbol\varepsilon\right)\right)/2$, $\mathrm{I}_{\boldsymbol\kappa}=\mathrm{tr}\left(\boldsymbol\kappa\right)$, and $\mathrm{I\!I}_{\boldsymbol\kappa}=\left(\left(\mathrm{tr}\boldsymbol\kappa\right)^2-\mathrm{tr}\left(\boldsymbol\kappa\cdot\boldsymbol\kappa\right)\right)/2$. It is crucial to note that the trace operation $\mathrm{tr}$ on any matrix, \emph{e.g.} $\mathbf{M}$, is here taken in a specific way that depends on the metric of the reference state in the following way $\mathrm{tr}\left(\mathbf{M}\right)\equiv\bar{a}^{\alpha\beta}M_{\alpha\beta}$. For homogenous and isotropic elastic shells, the total elastic energy takes the following form 
\begin{equation}
	\label{eq:Total_Energy}
		\mathcal{E}=\frac{Et}{2\left(1- \nu^2\right)}\int\mathrm{d}A\left[\left(1-\nu\right)\mathrm{tr}\left(\boldsymbol\varepsilon^2\right)+\nu\left(\mathrm{tr}\boldsymbol\varepsilon\right)^2+\frac{t^2}{12}\left(\left(1-\nu\right)\mathrm{tr}\left(\boldsymbol\kappa^2\right)+\nu\left(\mathrm{tr}\boldsymbol\kappa\right)^2\right)\right],
\end{equation}
where $E$ is the Young's modulus, $\nu$ the Poisson ratio, and $t$ the thickness of the shell. 
The terms in Eq.~\eqref{eq:Total_Energy} proportional to $t$ compose the stretching energy, whereas the ones proportional to $t^3$ the bending energy.

Symmetric stress and moment tensors are derived from Eq.~\eqref{eq:Total_Energy}, as conjugates of the in-plane and curvature strains, respectively. For simplicity, we choose the following linear constitutive laws for the components of stress, $\sigma^{\alpha\beta}$, and moment, $\mu^{\alpha\beta}$,
\begin{subequations}
	\label{eq:ConstitutiveLaw}
		\begin{align}
			\label{eq:StressStrain}
				\sigma^{\alpha\beta}&\equiv \frac{\delta\mathcal{E}}{\delta\varepsilon_{\alpha\beta}}\equiv\frac{Et}{1-\nu^2}\left[\left(1-\nu\right)\varepsilon^{\alpha\beta}+\nu\bar{a}^{\alpha\beta}\varepsilon^{\gamma}{}_{\gamma}\right]\\
			\label{eq:MomentCurvature}
				\mu^{\alpha\beta}&\equiv\frac{\delta\mathcal{E}}{\delta\kappa_{\alpha\beta}}\equiv\frac{Et^3}{12\left(1- \nu^2\right)}\left[\left(1-\nu\right)\kappa^{\alpha\beta}+\nu\bar{a}^{\alpha\beta}\kappa^{\gamma}{}_{\gamma}\right],
		\end{align}
\end{subequations}
where $\nu$ is the Poisson's ratio and $\delta\mathcal{E}/\delta f(x)$ stands for the functional derivative of Eq.~\eqref{eq:Total_Energy} with respect to $f(x)$.    

In the classical theory of shells~\cite{Koiter:1966,Koiter:1970,Fluegge:1973,Niordson:1985}, the principle of \emph{virtual work} allows us to derive the equations of equilibrium and boundary conditions. This principle is stated through the following relation
\begin{eqnarray}
	\label{eq:VirtualWork}
		\delta\mathcal{E}&=&\int\mathrm{d}A\left(\sigma^{\alpha\beta}\delta\varepsilon_{\alpha\beta}+\mu^{\alpha\beta}\delta\kappa_{\alpha\beta}\right)\nonumber\\
								  &+&\int\mathrm{d}A\left(f^\alpha\delta u_\alpha+p\,\delta v\right)+\oint\mathrm{d}s\left(T^\alpha\delta u_\alpha+m^\alpha\delta\Omega_\alpha+\mathcal{T}\,\delta v\right),
\end{eqnarray}
where $\delta\Omega^\alpha\equiv\mathrm{e}^{\alpha\beta}\left(b_{\beta\gamma}\delta u^\gamma+\partial_\beta\delta v\right)$ is the measure of rotation of a shell element and $\mathrm{e}^{\alpha\beta}$ is the permutation symbol ($\mathrm{e}^{12}=+1$, $\mathrm{e}^{21}=-1$, and $\mathrm{e}^{\alpha\beta}=0$ if $\alpha=\beta$). The second line in Eq.~\eqref{eq:VirtualWork} contains the following external forces and torques acting on the shell that balance with the stresses and moments:
\begin{itemize}
	\item Load $\mathbf{f}\mathrm{d}A$ that acts on an area element of the reference mid-surface:
	\begin{equation}
		\label{eq:ForceApplied}
			\mathbf{f}=f^\alpha\partial_\alpha\mathbf{S}+p\,\bar{\mathbf{N}}.
	\end{equation}
	\item Load $\mathbf{T}\mathrm{d}s$ that acts on a length element along the boundary of the reference mid-surface:
	\begin{equation}
		\label{eq:ForceBApplied}
			\mathbf{T}=T^\alpha\partial_\alpha\mathbf{S}+\mathcal{T}\,\bar{\mathbf{N}}.
	\end{equation}
	\item Moments acting on the boundary of the mid-surface are of two kinds, namely \emph{twisting moment} perpendicular to the boundary, $m_\perp$, and \emph{bending moment} parallel to the boundary,  $m_\parallel$:
	\begin{equation}
		\label{eq:MomentApplied}
			\mathbf{m}=m^\alpha\partial_\alpha\mathbf{S}=\left(m_\perp n^\alpha-m_\parallel\mathrm{l}^\alpha\right)\partial_\alpha\mathbf{S},
	\end{equation}
	where the directions $\hat{\mathbf{n}}$ and $\hat{\mathbf{l}}$ are defined as the perpendicular outward and tangent to the boundary edge, respectively. 
\end{itemize}
From Eq.~\eqref{eq:VirtualWork}, after carefully expressing the variations $\delta\varepsilon_{\alpha\beta}$ and $\delta\kappa_{\alpha\beta}$ through a fundamental variation of the embedding $\delta\mathbf{F}=\delta u^\alpha\partial_\alpha\mathbf{S}+\delta v\,\bar{\mathbf{N}}$ (details of this derivation is found in \cite{Niordson:1985}), we find the balance equations
\begin{subequations}
	\label{eq:Balance}
		\begin{align}
			\label{eq:Balance1}
				\bar{\nabla}_\alpha\sigma^{\alpha\beta}+2\bar{b}^\beta{}_\gamma\bar{\nabla}_\alpha\mu^{\gamma\alpha}+\mu^{\gamma\alpha}\bar{\nabla}_\alpha\bar{b}^\beta{}_\gamma+f^\beta&=0\\
			\label{eq:Balance2}
				\bar{\nabla}_\alpha\bar{\nabla}_\beta\mu^{\alpha\beta}-\bar{b}_{\alpha\gamma}\bar{b}^\gamma{}_\beta\mu^{\alpha\beta}-\bar{b}_{\alpha\beta}\sigma^{\alpha\beta}-p&=0
		\end{align}
\end{subequations}
and boundary conditions
\begin{subequations}
	\label{eq:Boundary}
		\begin{align}
			\label{eq:Boundary1}
				\left.\mu^{\alpha\beta} n_\alpha n_\beta\right|_{\partial\bar{\mathcal{S}}}&=m_\parallel\\
			\label{eq:Boundary2}
				\left.-\left(\bar{\nabla}_\alpha\mu^{\alpha\beta}\right) n_\beta-\partial_s\left(\mu^{\alpha\beta} n_\alpha\mathrm{l}_\beta\right)\right|_{\partial\mathcal{S}}&=\mathcal{T}-\partial_sm_\perp\equiv Q\\
			\label{eq:Boundary3}
				\left.\left[\sigma^{\alpha\beta}+\left(2\bar{b}^\alpha{}_\gamma-\bar{b}^\alpha{}_\rho n^\rho n_\gamma\right)\mu^{\gamma\beta}\right] n_\beta\right|_{\partial\bar{\mathcal{S}}}&=T^\alpha+b^\alpha{}_\beta\mathrm{l}^\beta m_\perp\equiv M^\alpha,
		\end{align}
\end{subequations}
where $Q$ is an effective shear force and $M^\alpha$ is an effective membrane force component.
From this point on we assume that the body surface forces are zero, $f^\alpha=p=0$. 

\subsection{Scaling analysis}
\label{suppsec:subsec:Scaling analysis}

We first discuss scaling laws for a flat plate of length $L$, width $w$, and thickness $t$, Young's modulus $E$, and Poisson ratio $\nu$. The in-plane stretching may be neglected for a flat plate under purely bending loads \cite{Landau1959aa}. This neglect can be justified as the simplification of \eqref{eq:StrainSmall} to $\varepsilon_{\alpha\beta}=(1/2)\left(\partial_\alpha u_\beta+\partial_\beta u_\alpha\right)+\partial_\alpha v\partial_\beta v$. In this case, the deformations are forced by out-of-plane bending, and therefore, if $v$ scales as the applied displacement $\delta$, then $u_\beta$ scales quadratically with $\delta$ as $\delta^2/L$ and  $\varepsilon_{\alpha\beta}$ as $\delta^2/L^2$. In addition, the curvature in this case scales linearly with $\delta$ as $|b_{\alpha\beta}|=|\partial_\alpha\partial_\beta v|\sim \delta/L^2$.

The total elastic energy $\mathcal{E}$ for deforming a plate is given by the sum of two terms, stretching and bending energies. The bending energy scales as $Et^3Lw/12(1-\nu^2) \times \delta^2/L^4$, while the stretching energy can be neglected because it scales a factor $(\delta/L)^2$ smaller as $EtLw/(1-\nu^2) \times \delta^4/L^4$.
Under this assumption, which decouples the stretching from the bending, we may estimate the stiffness of a plate when a shear vertical force, $F$, is applied at its edge. Adding the work done by the shear vertical force, $F\delta$, in order to deflect the plate by an amount $\delta$, energy minimization yields the following scaling for the deflection: $v\sim 4FL^3(1-\nu^2)/\left(t^3wE\right)$. The stiffness of the plate is 
\begin{equation}
   K_{\mbox{\small plate}}=\frac{F}{\delta} = \frac{Ewt^3}{4 L^3(1-\nu^2)}.
\end{equation}

We next consider a curved shell, where the stretching energy can no longer be neglected. Let $\ell$ denote the characteristic length scale of the deformation resulting purely from an external force to bend the shell out of plane. Then the scales of in-plane and curvature strains that establish in response to this forcing are $\delta/R$ and $\delta/\ell^2$ respectively. This yields, according to the linear constitutive laws in Eqs.~\eqref{eq:ConstitutiveLaw}, the following scaling for the stress and moment components: $\sigma^{\alpha\beta} \sim \mathcal{O} \left[ E t \delta/R \right]$ and $\mu^{\alpha\beta}\sim \mathcal{O} \left[Et^3 \delta/\ell^2 \right]$, where we let $R$ be the smallest principal radius of curvature. If we compare the orders of the terms in Eq.~(\ref{eq:Balance1}-\ref{eq:Balance2}), we have
\begin{equation}
	\label{eq:Scaling1}
		\bar{\nabla}_\alpha\sigma^{\alpha\beta}\sim\mathcal{O}\left[\frac{Et\delta}{R \ell}\right],\quad 2\bar{b}^\beta{}_\gamma\bar{\nabla}_\alpha\mu^{\gamma\alpha}\sim\mu^{\gamma\alpha}\bar{\nabla}_\alpha\bar{b}^\beta{}_\gamma \sim\mathcal{O}\left[\frac{Et^3\delta}{R\ell^3}\right],
\end{equation}
and the ratio of the latter terms to the former is
\begin{equation}
	\label{eq:Ratio1}
		r \sim\mathcal{O}\left[\frac{t^2}{\ell^2}\right].
\end{equation}
Similarly, for Eq.~\eqref{eq:Balance2} we have that
\begin{equation}
	\label{eq:Scaling2}
		\bar{b}_{\alpha\beta}\sigma^{\alpha\beta}\sim\mathcal{O}\left[\frac{Et\delta}{R^2}\right] \quad
		\bar{b}_{\alpha\gamma}\bar{b}^\gamma{}_\beta\mu^{\alpha\beta}\sim\mathcal{O}\left[\frac{Et^3\delta}{R^2\ell^2}\right],
\end{equation}
and the ratio of the latter to the former is also $r$.

The length scale over which a shell deforms is commonly much larger than the thickness of the shell, implying $r \ll 1$, and therefore we can approximate Eq.~(\ref{eq:Balance1}-\ref{eq:Balance2}) by
\begin{subequations}
\begin{align}
	\label{eq:Balance1Approx}
		\bar{\nabla}_\alpha\sigma^{\alpha\beta}=0, \\
	\label{eq:Balance2Approx}
		\bar{\nabla}_\alpha\bar{\nabla}_\beta\mu^{\alpha\beta}-\bar{b}_{\alpha\beta}\sigma^{\alpha\beta}=0.
\end{align}
\end{subequations}

Balancing the first term in \eqref{eq:Balance2Approx}, which scales as $\mathcal{O}\left[ Et^3 \delta/\ell^4 \right]$, with the second term yields
\begin{align}
 \ell = \sqrt{Rt}.
 \label{eq:LengthScale}
\end{align}

The applied shear stress $Q$ in Eq.~\eqref{eq:Boundary2} scales as $\mathcal{O}\left[ Et^3\delta/\ell^3 \right]$.
The stiffness of the shell, defined as the total force applied on that edge divided by the displacement, scales as 
\begin{align}
K\sim \mathcal{O}\left[ \dfrac{Et^3 w}{\ell^3} \right] \sim \mathcal{O}\left[ Ew \left(\dfrac{t}{R}\right)^{3/2} \right].
\label{eq:StiffnessScale}
\end{align}
Eqs.~(\ref{eq:LengthScale}-\ref{eq:StiffnessScale}) are the main results of the scaling analysis verified using computations and experiments on a thin shell.

\section{Experimental measurement of stiffness of shells}
\label{suppsec:experimental shells}

\subsection{Fabrication of Arches}
\label{suppsec:subsec:fabrication of arches}
Fabrication of arches started with preparation of a Stereolithography (STL) file of the arch mould on a computer aided design (CAD) software tool (Solidworks, Dassault Systemes). The arch length ($L$), its width ($w$), thickness ($t$), and the longitudinal ($R_L$) and transversal ($R_T$) radii of curvature formed the relevant input parameters to the STL file. This STL file, in turn formed the input to the 3D printer (ProJet 460Plus, 3D Systems) to print the relevant mould for the arch. A component infiltrate (Colorbond) was applied to impart additional strength to the part printed with a powder composite (Visijet PXL Core) combined with a binding agent (Visijet PXL). The powder composite based printing method was adopted for reasons of convenience. It offered fast printing with reasonable layer resolution of 100 micrometers, while maintaining minimal corrugated step features that are inevitable when printing curved objects. Furthermore, with a melting point of 1450 $^{\circ}$C, powder composite based moulds proved very stable when baking the elastomer based arches as discussed in the following.

The printed mould served as a shell, few millimeters in thickness, with one side left open. Poly (dimethyl siloxane) (PDMS) silicone elastomer (Sylgard 184, Dow Corning) was employed to cast the arch in the mould. The PDMS base polymer and the curing agent were thoroughly mixed in a centrifuge (Thinky ARE-310). Since the ratio of base polymer to curing agent controls the elastic modulus for PDMS, the same ratio of 5 parts base polymer to 1 part of curing agent by weight was consistently maintained across all fabricated arches employed in experiments discussed here. This mixture was poured into the printed mould and degassed for a 30 - 40 minute duration, to remove all trapped air bubbles. The degassed PDMS mixture in the arch mould was then transferred to an oven and baked at 75 $^{\circ}$C for 12 hours to obtain the experimental arches as shown in fig.~\ref{fig extended:experiments}a. 

During an experiment, the fabricated arch was mounted on the experimental rig with clamps that were custom fabricated to exactly match the arch curvature. The clamps were 3D printed (Stratasys Dimension 1200es)  with ABSPlus (Acrylonitrile butadiene styrene) thermoplastic material (glass transition temperature 108$^{\circ}$C). The arch clamps were designed such that the clamp spacing exactly matched the arch thickness to ensure the arch shape (cross-sectional area and radius of curvature) remained unchanged when tightening the clamps. Glue was applied between the arch and clamp to maintain a strict no-slip condition during loading.

\subsection{Experimental setup and protocol}
\label{suppsec:subsec:Experimental setup and protocol}
The arch with clamps was mounted on to the experimental rig (fig.~\ref{fig extended:experiments}a). One end of the clamped arch was affixed to a rigid plate attached to a free-moving horizontal translation stage to maintain constant horizontal distance between the clamps during the load test. This prevented the arch from changing its length under loading. The other side of the clamped arch was connected to a force sensor (SMT-1 S-Type Load Cell, Interface Advanced Force Measurement) affixed to a translation stage that moved along the vertical direction. The force sensor was interfaced with a data acquisition system (LabView, National Instruments) to obtain the force measurements from the load test. 

Unlike computer simulations, experimental arches must contend with gravitational force owing to mass of the arch sample. Circumvention of gravitational effects therefore requires that arch samples cannot be too long or too thick, lest they slacken under their own weight in the central region between the two end clamps. Arch slackening causes a systematic error in the experimental determination of bending stiffness $K$, for then the initial displacement must first straighten the arch, and only then proceed to load it. The arches were therefore designed to span the range of values for $L^2/Rt$ through a combination of the fabrication parameters, while ensuring the length $L$ and thickness $t$ stayed within reasonable bounds to avoid slackening.

Prior to each experiment, the vertical translation stage was positioned to ensure the arch sample was perfectly horizontal. This position was set as the zero displacement point for the vertical translation stage. An initial force reading was then taken to measure the force offset value arising from weight of the clamps and the rigid metallic connector between the clamp and force sensor. This offset was then subtracted from the measured force.

The load test was performed under quasi-static conditions of the arch sample by providing small displacements (quasi-static steps) of $5 \times 10^{-5}$ m ($50~\mu$m) per step for a total of 10 quasi-static steps ($5 \times 10^{-4}$ m or $500~\mu$m). A force measurement was conducted after each quasi-static displacement with the LabView interface at 2 KHz sampling frequency for 1 second duration. The arithmetic mean over the 2000 force data points was recorded as the measured force for a given displacement. The slope of the force-displacement curve obtained from a linear fit of experimental data then provided the bending stiffness $K$ for the arch sample. Three experimental runs were conducted for each arch and their force-displacement curves were measured to be reproducible to within measurement error.

\subsection{Calibration of PDMS material constants}
\label{suppsec:subsec:Calibration of PDMS material constants}
Unlike crystalline solids \cite{Kittel:2004}, there exists no microscopic theory that permits calculation of material constants for amorphous media which include polymers, alloys, granular solids etc. Whereas part of the difficulty arises from the structural disorder which imparts an effective modulus to amorphous solids, material constants also sensitively depend upon the preparation protocol. In particular, the elastic constants and poisson ratio of PDMS based soft solids depend upon the base polymer to cross-linker ratio, curing temperature and baking time. This calls for calibration of material constants of PDMS solids used for curved shells in this study.

Flat PDMS test specimens were prepared using same preparation protocols as the PDMS curved shells used in this study, viz. 5:1 base to curing agent ratio was thoroughly mixed in a centrifuge (Thinky ARE-310), followed by removal of trapped air bubbles via degassing for 30 - 40 minutes, and then baked at 75 $^{\circ}$C for 12 hours. The dimensions of the PDMS calibration specimens were $7 \times 10^{-2}$ m (length), $5 \times 10^{-2}$ m (width), and $5 \times 10^{-3}$, $7 \times 10^{-3}$, and $9 \times 10^{-3}$ m (thickness) in order to match the dimensions of curved PDMS shells used in experiments.

\begin{figure}
\centering
\includegraphics{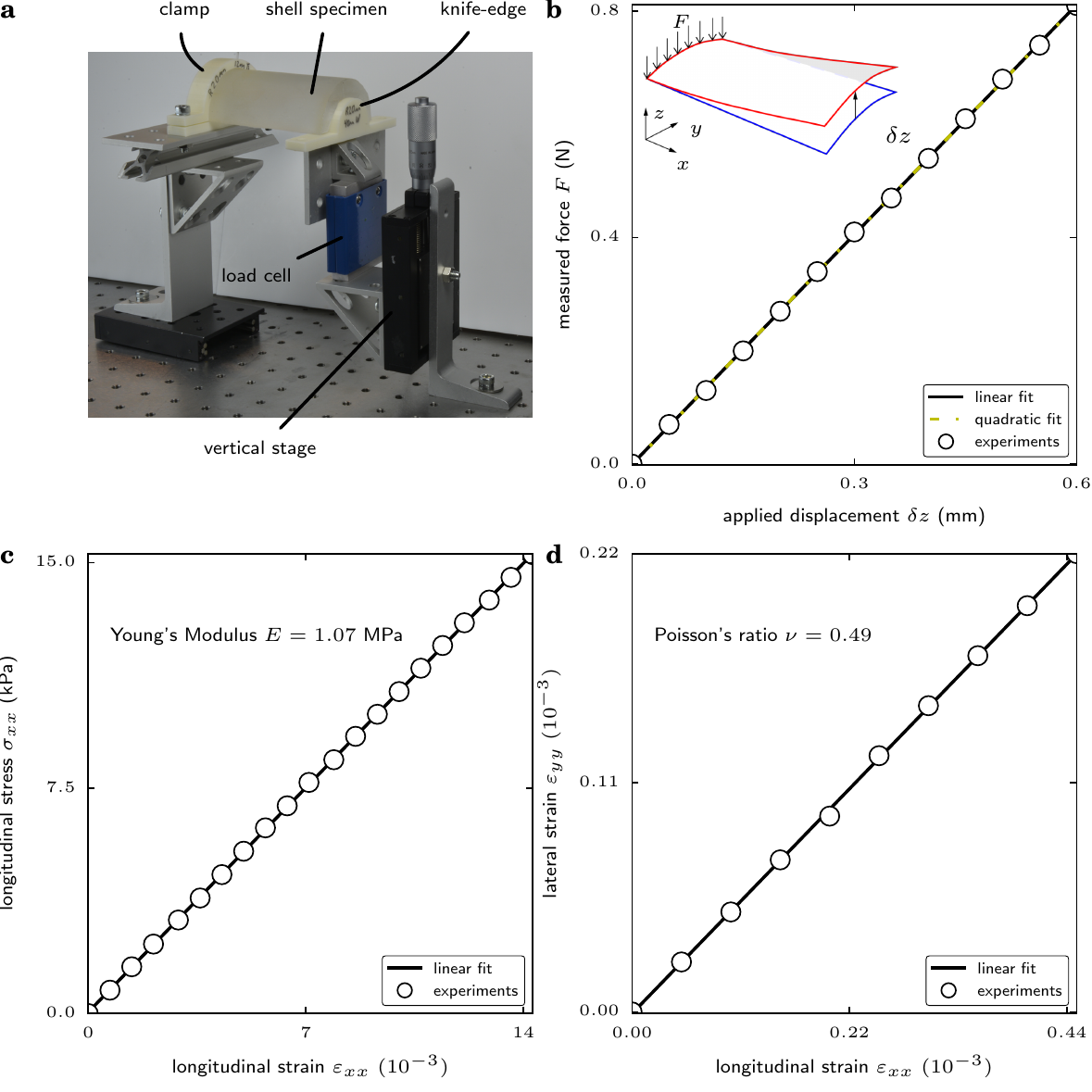}
\caption{Experimental measurement of stiffness of arched shells. {\bf a.} The experimental setup used in stiffness measurements. {\bf b.} Representative data that show linearity of force-displacement curve. The best fit quadratic is indistinguishable from the linear fit to within sensor resolution. {\bf c.} The Young's modulus and {\bf d.} Poisson's ratio of the PDMS material used in fabricating the shells were estimated from simultaneous stress and strain measurements during an extension test of a rectangular PDMS block.}
\label{fig extended:experiments}
\end{figure}

The fabricated calibration specimens were mounted onto the experimental rig using a clamping mechanism (3D printed from ABS plastic) as detailed in section~\ref{suppsec:subsec:Experimental setup and protocol}. Clamps on either side of the calibration specimen were mounted onto the experimental rig with one minor change in the setup. Instead of vertical loading perpendicular to horizontal plane of the flat sheet, a horizontal loading in plane along the direction of longitudinal axis of the calibration sheet was performed. The tensile loading yielded a simple stress-strain curve (see Fig.~\ref{fig extended:experiments}c,d). The Young's modulus obtained from a linear fit (Fig.~\ref{fig extended:experiments}c) through the experimental data was $1.07 \times 10^{6}$ N/m$^2$ for calibration sheets of thickness $5 \times 10^{-3}$~m.

Whereas the same calibration specimens were used, the experimental setup was further modified for measurement of Poisson ratio ($\nu$). A camera (Pixelink PL-B778, 5 Megapixels) with a macro lens (single focal length 106 mm) was mounted atop the experimental setup to record the lateral strain $\epsilon_{\perp}$ acting normal to the direction of the applied tensile loading. A longitudinal strain $\epsilon_{||}$ (absolute strain of $1 \times 10^{-5}$ m) was administered and the lateral strain ($\epsilon_{\perp}$) obtained from processing the images. The slope obtained from a linear fit (Fig.~\ref{fig extended:experiments}d) of the  lateral strain ($\epsilon_{\perp}$) versus longitudinal strain ($\epsilon_{||}$) curve yielded a poisson ratio of $\nu = $0.48, 0.47 and 0.49 for calibration specimens with thickness $5 \times 10^{-3}$, $7 \times 10^{-3}$, and $9 \times 10^{-3}$ m respectively.

\subsection{Data Analysis: Determination of Bending Stiffness $K$}
A typical force-displacement curve from an experimental load test of the arch is shown in Fig.~\ref{fig extended:experiments}b. The data collected in discrete steps of $5 \times 10^{-4}$ m ($500~\mu$m) were fit to a continuous curve in order to ascertain the arch bending stiffness $K$ from the slope of the force-displacement curve. In the shell theory formulation used in our predictions, linearization of the kinematic relations in equations~\eqref{eq:DefMeasure}, as in equations~\eqref{eq:DefMeasureSmall}, and constitutive laws in equations~\eqref{eq:ConstitutiveLaw} are assumed. Therefore, the theoretically predicted scalings arise within the linear response regime for force-displacement curves. The loading experiments therefore had to be conducted within the linear response regime. A verification of this requirement was performed using polynomial fits of the experimental force-displacement curves. To this end, polynomial fits to linear (solid line in Fig.~\ref{fig extended:experiments}b) as well as quadratic (dashed curve in Fig.~\ref{fig extended:experiments}b) order were performed. The linear fit yielded $F = -4.4 \times 10^{-4} + 1.4 \times 10^3x$ and the quadratic fit yielded $F = -1.2 \times 10^{-3} +1.4 \times 10^3x - 1.4 \times 10^{4} x^2$. The zeroth order coefficients for both linear and quadratic fits being below the force sensor noise floor ($1 \times 10^{-2}$ N), they are treated equivalent to zero. The two fits are in agreement at linear order, with very slow variation in quadratic coefficient with displacement $x$. Hence, the experiments are considered to be operating within the linear response regime. The bending stiffness $K$ is computed from the slope of force-displacement curves using linear fits to experimental data.

\section{Biological feet}
\label{suppsec:Biological feet}
\subsection{Validity of the shell approximation}
\label{suppsec:subsec:Validity of the shell approximation}
The shell may be decomposed in two parts: a soft part of length scale $\ell$ near the loaded edge as described by the scaling analysis in section~\ref{suppsec:subsec:Scaling analysis}, and the rest of the shell of length $L'$ which acts as an Euler beam. 
The approximation of the shell as an Euler beam  is supported by the results of the computation (Fig.~\ref{fig extended:numerics}d, e) of the distribution of in-plane longitudinal stress $\sigma_{xx}$ throughout the shell.
The neutral plane for the cross-section of a shallow shell whose central plane is approximated by $z=y^2/2R_T$ may be determined as 
\begin{align}
 z_n = \dfrac{1}{w} \int_{-w/2}^{w/2} \dfrac{y^2}{2R_T} dy = \dfrac{w^2}{24 R_T},
\end{align}
which is plotted in figure \ref{fig extended:numerics}(c).
This expression for the neutral plane agrees with the zero-contour of the longitudinal stress, as shown in Figure \ref{fig extended:numerics}(e). 
The second moment of the cross section of such a shell is
\begin{align}
 I_n = \int_{-w/2}^{w/2} \left( \dfrac{y^2}{2R_T} - z_n \right)^2~t~ dy = \dfrac{w^5t}{720 R_T^2}.
\end{align}
The bending rigidity of the corresponding Euler beam $B_n = \dfrac{Etw^5}{720 R_T^2}$, and the contribution to the stiffness of the shell is $\dfrac{3B_n}{L'^3} = \dfrac{Etw^5}{240 R_T^2 L'^3}$.

This Euler beam model loses validity near the loaded edge.
In this region, which we call the edge-boundary layer, the deformation departs from the simple pattern required for the Euler beam approximation. 
The scaling estimates we derived in the previous subsection hold in this region.
Based on those scaling estimates, the stiffness contributed by this region to the shell scales as $K$ calculated in equation~\eqref{eq:StiffnessScale} and occurs in series with the stiffness of the Euler beam.
Since the softest spring determines the stiffness of springs in series, the edge-boundary layer determines the elastic response of the shell if
\begin{align}
Ew \left(\dfrac{t}{R_T}\right)^{3/2} \ll \dfrac{Etw^5}{R_T^2 L'^3} \qquad \text{ or equivalently, } \qquad \eta \equiv \dfrac{\ell L'^3}{w^4} \ll 1. 
\label{eqn:Ratio}
\end{align}
To estimate $L'$, we subtract $2\ell$ each to account for the edge-boundary layers at the loaded and the clamped ends, implying $L'=L-4\ell$. 
If the criterion in (\ref{eqn:Ratio}) is satisfied, the bending deformation is localized to the edge-boundary layer.

Using Monte Carlo simulations, as described in the methods, we determine the distribution of $\eta$ in human feet.
The first, second and third quartiles for the distribution of $\eta$ are 0.0033, 0.030, and 0.14 respectively.
The 95$^\text{th}$ percentile for the distribution is 0.65, based on which we conclude that for the vast majority of feet, $\eta \ll 1$.

\renewcommand{\refname}{References for the appendix}

\end{document}